\documentclass[a4paper,11pt]{article}
\usepackage{aaskaiid}
\usepackage{subcaption} 
\usepackage{orcidlink}

\usepackage{hyperref}
\usepackage[nameinlink,capitalise]{cleveref}
\usepackage{amsfonts,amsmath,bm,mathtools}
\usepackage{ulem,comment,siunitx}
\DeclareSIUnit\Jy{Jy}
\DeclareSIUnit\parsec{pc}
\usepackage{tikz}
\usetikzlibrary{arrows.meta}

\newcommand{\planck}{{\it Planck}\ }
\newcommand{\bk}{{\boldsymbol{k} }}

\newcommand{\hi}{\textrm{H\textsc{i}}}
\newcommand{\fnl}{f_\mathrm{NL}}
\title{Beyond $\Lambda$CDM with the SKA Observatory -- II\\
Unveiling the Secrets of the Early Universe}
\ShortTitle{Beyond $\varLambda$CDM with the SKAO -- II: Tests of Inflation}

\author[1,2,3]{José Fonseca\orcidlink{0000-0003-0549-1614}}
\ShortName{Fonseca et al.} 
\author[4,5,6]{Benedict Bahr-Kalus\orcidlink{0000-0002-4578-4019}}
\author[7,8,9,3]{Mario Ballardini\orcidlink{0000-0003-4481-3559}}
\author[4,10,11,12]{Matilde Barberi-Squarotti\orcidlink{0009-0007-8964-5807}}
\author[13,14]{Steven Cunnington\orcidlink{0000-0001-6594-107X}}
\author[3]{S\^ecloka L. Guedezounme\orcidlink{0009-0000-9200-8584}}
\author[7,8,19,3]{Dionysios Karagiannis\orcidlink{0000-0002-4927-0816}}
\author[4,5]{Samantha J. Rossiter\orcidlink{0009-0005-7953-2396}}
\author[15,16,17]{Ziad Sakr\orcidlink{0000-0002-4823-3757}}
\author[18]{Cora Uhlemann\orcidlink{0000-0001-7831-1579}}
\author[4,5,6,3]{Stefano Camera\orcidlink{0000-0003-3399-3574}}
\author[3]{Bikash R. Dinda\orcidlink{0000-0001-5432-667X}}
\author[19,20]{Cláudio Gomes\orcidlink{0000-0001-6022-459X}}
\author[3]{Roy Maartens\orcidlink{0000-0001-9050-5894}}
\author[3,21]{Mario G. Santos\orcidlink{0000-0003-3892-3073}}

\affiliation[1]{Instituto de Astrof\'isica e Ci\^encias do Espa\c{c}o, Universidade do Porto, CAUP, Rua das Estrelas, PT4150-762 Porto, Portugal}
\emailAdd{jose.fonseca@astro.up.pt}
\affiliation[2]{Departamento de F\'isica e Astronomia, Faculdade de Ci\^{e}ncias, Universidade do Porto, Rua do Campo Alegre 687, 4169-007 Porto, Portugal}
\affiliation[3]{Department of Physics and Astronomy, University of the Western
Cape, Robert Sobukwe Road, Cape Town 7535, South Africa}
\affiliation[4]{Dipartimento di Fisica, Universit\`a degli Studi di Torino, Via P.\ Giuria 1, 10125 Torino, Italy}
\affiliation[5]{INAF -- Istituto Nazionale di Astrofisica, Osservatorio Astrofisico di Torino, Strada Osservatorio 20, 10025 Pino Torinese, Italy}
\affiliation[6]{INFN-- Istituto Nazionale di Fisica Nucleare, Sezione di Torino, Via P.\ Giuria 1, 10125 Torino, Italy}
\affiliation[7]{Dipartimento di Fisica e Scienze della Terra, Universit\`a degli Studi
di Ferrara, Via G.\ Saragat 1, 44122 Ferrara, Italy}
\affiliation[8]{INFN-- Istituto Nazionale di Fisica Nucleare, Sezione di Ferrara, via Giuseppe Saragat 1, 44122 Ferrara, Italy}
\affiliation[9]{INAF/OAS Bologna, via Piero Gobetti 101, 40129 Bologna, Italy}

\affiliation[10]{Dipartimento di Fisica, Universit\`a degli Studi di Milano, via G.\ Celoria 16, 20133 Milano, Italy}
\affiliation[11]{INFN -- Istituto Nazionale di Fisica Nucleare, Sezione di Milano, via G.\ Celoria 16, 20133 Milano, Italy}
\affiliation[12]{INAF -- Istituto Nazionale di Astrofisica, Osservatorio Astrofisico di Brera-Merate, via Brera 28, 20121 Milano, Italy}

\affiliation[13]{Institute of Cosmology \& Gravitation, University of Portsmouth, Dennis Sciama Building, Portsmouth, PO1 3FX, UK}
\affiliation[14]{Jodrell Bank Centre for Astrophysics, Department of Physics \& Astronomy, The University of Manchester, Manchester M13 9PL, UK}

\affiliation[15]{Instituto de Física Teórica UAM-CSIC, Campus de Cantoblanco, 28049 Madrid, Spain}
\affiliation[16]{Institut de Recherche en Astrophysique et Plan\'etologie (IRAP), Universit\'e de Toulouse, CNRS, UPS, CNES, 14 Av. Edouard Belin, 31400 Toulouse, France}
\affiliation[17]{Universit\'e St Joseph; Faculty of Sciences, Beirut, BP-11514, Lebanon}
\affiliation[18]{Fakultät für Physik, Universität Bielefeld, Postfach 100131, 33501 Bielefeld, Germany}
\affiliation[19]{Van Swinderen Institute, University of Groningen, Nijenborgh 3, 9747 AG Groningen, The Netherlands}

\affiliation[19]{Centro de Física das Universidades do Minho e do Porto, Faculdade de Ciências da Universidade do Porto, Rua do Campo Alegre s/n, 4169-007 Porto, Portugal}
\affiliation[20]{OKEANOS, Universidade dos Açores, Rua Professor Doutor Frederico Machado 4, 9900-140 Horta, Portugal}
\affiliation[21]{South African Radio Astronomy Observatory (SARAO), Liesbeek House, River Park, Gloucester Road, Mowbray, Cape Town, 7700, South Africa}

\abstract{The origins of the universe remain one of the biggest mysteries in modern cosmology. While the Planck satellite has provided a wealth of information about the early universe, there is still much to be discovered. The Square Kilometre Array Observatory (SKAO) offers a unique opportunity to probe the universe's infancy, going beyond the current limitations of our knowledge. By measuring the power spectrum of biased tracers of the dark matter distribution on the largest cosmological scales and exploring beyond 2-point statistics, SKAO will enable us to refine our understanding of the primordial universe, including the shape of the inflationary power spectrum and the presence of primordial non-Gaussianity. In this chapter we will review recent works looking at the potential of SKAO's surveys, and how synergies with other surveys can revolutionize our understanding of the origins of the cosmos.
}

\begin{document}
\maketitle

\section{Introduction}

The primordial universe is a fascinating period of the history of the Cosmos despite our lack of understanding of the times prior to the Big Bang nucleosynthesis. The best proposal and framework of ideas to describe the very early universe is the so-called cosmic inflation \citep{Starobinsky:1980te,Guth:1980zm,Sato:1980yn,Linde:1981mu,Albrecht:1982wi,Hawking:1982ga,Linde:1983gd}, which corresponds to a period of nearly an exponential expansion. Such an epoch would naturally solve conceptual problems with the hot big bang model, namely the horizon, flatness and monopole problems (citations). It was then understood that the local quantum fluctuations in the quantum field(s) driving the expansion were stretched and amplified, providing the seeds for the observed large-scale structure (LSS) of the Universe \citep{Mukhanov:1981xt}. 

The most successful probe to constrain inflation has been the \planck satellite \citep{2020A&A...641A..10P} with its observation of the Cosmic Microwave Background (CMB). Its findings are consistent (broadly as a general framework) with the simplest model of inflation where inflation is governed by a single slowly rolling scalar field, called the inflaton. Single-field slow-roll inflation is the scenario that best describes CMB observations, and is the one statistically preferred 
by the data. Still, there is room 
for more complex inflationary scenarios. For simplicity, \planck assumes the neutrino sector fixed to one massive neutrino with a minimal mass and two massless neutrinos. This is justified as \planck sensitivity could not go beyond these assumptions and such a choice made computation time more efficient. If one opens up the possibility for the neutrino masses and the effective number of relativistic degrees of freedom, $N_{\rm eff}$, to be free parameters, then \planck predicts slightly lower values for the spectral scalar index, $n_{\rm s}$ \citep{Gerbino:2016sgw,Gomes:2018uhv}. Recent data releases from the Atacama Cosmology Telescope (ACT DR6) and baryonic acoustic oscillation data from the Dark Energy Spectroscopic Instrument (DESI DR2) combined with \planck and BICEP/Keck 2018 (BK18) predicts a slightly lower value for the scalar-to-tensor ratio, $r<0.038$, and a higher one for the spectral scalar index, namely $n_{\rm s}=0.9743\pm0.0034$ \citep{ACT:2025fju,ACT:2025tim,DESI:2024mwx,DESI:2024uvr}. This motivates to seek more complex models, or include higher-order corrections from a putative UV-completion of gravity, as well as more observables that can probe the very early universe. While \planck provided tight constraints, it could not go beyond the linear approach and definitively distinguish between different inflationary scenarios or rule out more complex models.

One of the simplest extensions to the base model is to consider that the variance of the quantum field fluctuations i.e., the primordial power spectrum is not a single power law as a function of the cosmological scale, specifically the wavenumber $k$. The running of the power spectrum (see below) is highly suppressed in single-field slow-roll inflation. Hence providing a tighter measurement of such a parameter can give further consistency to cosmic inflation. While in the simplest inflationary model the primordial fluctuations follow (almost) a random Gaussian distribution \citep{barrow1990}, other models predict deviations from this behaviour \citep{bartolo2004}. This means that the odd correlation functions of the primordial gravitational potential created by the fluctuations in the energy density of the inflaton (or the field responsible for the structure in the universe) are no longer zero. Later on in the history of the universe, such a deviation from Gaussianity also affects how matter clusters in the gravitational potentials giving rise to a scale-dependent galaxy bias of
dark matter tracers, which is relevant on very large scales \citep{Matarrese:2008nc,Dalal:2007cu}. Additionally, the primordial power spectrum on very large scales can carry information from temporary violations of the slow-roll conditions. Different extended models predict that the shape of the power spectrum in such scales to be substantially different from the standard scenario. 

The search for subtle deviations from the standard picture is crucial as these deviations, if found, would provide invaluable clues about the specific physical processes that occurred during the inflationary epoch. Hence our focus here on three key observables that are sensitive to the fine details of inflation: the running of the spectral index, primordial features (arising from temporary violations of slow-roll), and primordial non-Gaussianity. 

Hunting for such effects will only be possible with very large-scale cosmological surveys as the ones planned with the Square Kilometre Array Observatory (SKAO). The {\it Wide Band 1 Survey} covering
$20,000 \deg^2$ with an integration time of approximately $10,000$ hours in the sky will provide a wide continuum
galaxy survey with a number density of sources of $n\simeq 1.4\rm arcmin^{-2}$, and \hi\ intensity mapping (IM) survey in the redshift
range $z = 0.35 - 3$ \citep{Bacon:2018dui}. For more details on the specification of each survey please look at the respective SKAO science chapter in Radio Continuum Galaxy survey \citep{Asorey01.2026.SKA} and \hi\ IM survey \citep{Wolz01.2026.SKA}. Such large cosmological surveys will be complementary to future Stage-V spectroscopic galaxy surveys in the optical and infrared such as MUST \citep{2024arXiv241107970Z} or MegaMapper \citep{2022arXiv220904322S} -- see also \citet{2022arXiv220903585S} for other proposals. Their complementarity is not just from observing more objects, but fundamentally from the fact that \hi\ IM probes the lighter halos while spectroscopic surveys look at the massive side of the halo-mass function \citep{2020JCAP...09..054V}.




\section{The primordial universe and the prospects to test it with the SKAO}

\subsection{Running of the spectral index}


In the simplest realisation of inflation, the primordial scalar perturbations have an almost scale-invariant power spectrum 
\begin{equation}
    P_{\rm s}(k) = A_{\rm s} \left(\frac{k}{k_0}\right)^{n_{\rm s} -1} \, .
    \label{eq:scalar_pow}
\end{equation}
Here, $A_{\rm s}$ is the scalar amplitude, and $n_{\rm s}$ is the scalar spectral index. This index quantifies the "tilt" of the spectrum, with $n_{\rm s} = 1$ representing perfect scale-invariance. \planck has tightly constrained this value to $n_{\rm s} = 0.9649 \pm 0.0042$ (at 68\% CL, corresponding to \planck TT, TE, EE + lowE + lensing), which is consistent with the predictions of simple slow-roll inflation. The pivot scale, $k_0$, is a reference wave number conventionally set to $k_0=0.05\;\mathrm{Mpc}^{-1}$.
 
However, constraints on the duration of inflation together with the BICEP3/Keck bounds on the gravitational wave background put pressure on this realisation \citep{2022PhRvD.106f1301E}. To reconcile this tension, one might need to invoke exotic reheating mechanisms, multi-field inflation, or, to preserve single-field slow-roll inflation, the presence of non-trivial higher-order derivatives of the inflaton potential. 

In the latter case, the spectral tilt itself gains a scale-dependence different from the simplest single-field slow-roll model. We call this the running, i.e., the scale-dependence of the spectral index. This is a key observable for distinguishing between different inflationary models. In this case, the scalar power spectrum takes the form:
\begin{equation}
    P_{\rm s}(k) = A_{\rm s} \left(\frac{k}{k_0}\right)^{n_{\rm s} -1 + \frac12 \alpha_{\rm{s}}\ln(k/k_0) + \ldots} \,,
    \label{eq:scalar_pow_alpha}
\end{equation}
with a running parameter $\alpha_{\rm s}$ \citep{Kosowsky:1995aa}. This parameter is very small and  well inside  current observational bound of $\alpha_{\rm{s}} =  -0.0045 \pm 0.0067$ at 68\% CL, corresponding to \textit{Planck} TT, TE, EE + lowE + lensing \citep{2020A&A...641A..10P}. In a very broad range of inflationary models and alternative scenarios we expect $10^{-4} \lesssim |\alpha_{\rm s}| \lesssim 10^{-3}$  \citep{Adshead:2010mc,Lehners:2015mra,2022PhRvD.106f1301E,Martin:2024nlo}. 

\begin{figure*}
    \centering    
\includegraphics[width=0.45\textwidth]{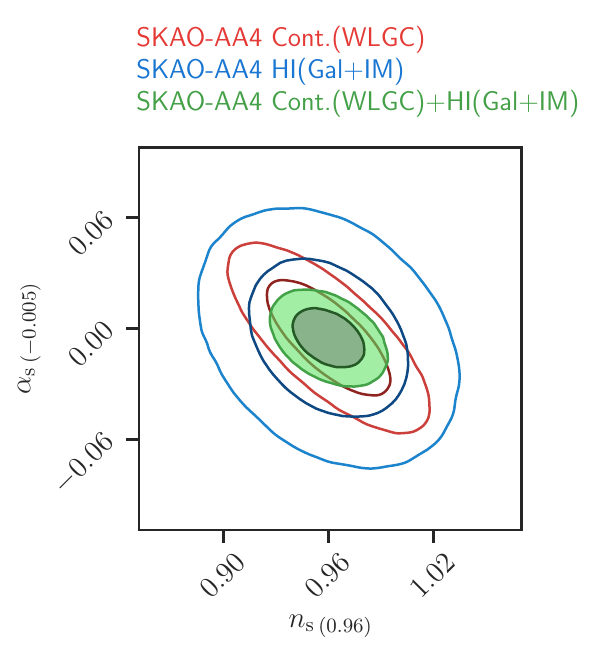} 
\includegraphics[width=0.485\textwidth]{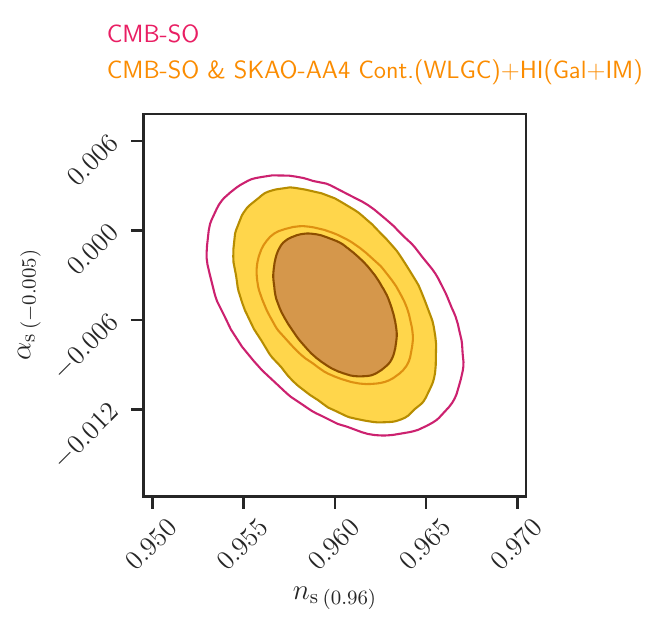}    
\caption{(Left): Forecast 1- and 2-$\sigma$ contours on the spectral index $n_{\rm s}$ and its running parameter $\alpha_{\rm s}$ using SKAO continuum or \hi ~probes and their combination following the AA4 specifications. (Right): Forecast 1- and 2-$\sigma$ contours on the same parameters using SKAO probes for the same SKAO configuration combined with a CMB future experiment Simons Observatory-like (CMB-SO) survey in comparison to contours obtained from CMB-SO alone. Here 'WLGC' stands for Weak Lensing (WL) shear correlations combined with Galaxy Clustering (GC) correlations and their cross-correlations; while 'Gal+IM' means correlations of Galaxies (Gal), with redshift measured from the 21 cm hyperline in Hydrogen atom, combined with the Intensity Mapping (IM) probe.}
    \label{fig:nsrun_SKAO}
\end{figure*}

We performed a forecast on $\alpha_{\rm s}$ shown in Fig.~\ref{fig:nsrun_SKAO} following the method and settings of \cite{Casas:2022vik}. We additionally account for the non-linearities for the continuum angular correlations probes following \cite{Euclid:2025hlc} while we adopt conservative settings --- $\ell_{\rm max}\sim 1000$ and $k_{\rm max} \sim 3 \,h\,{\rm Mpc}^{-1}$. This keeps the agreement between {\tt HMCode} \cite{Mead:2015yca}, which we used, and {\tt NGenHalofit} \citep{Smith:2018zcj} within two percent. The non-linear modelling of \cite{Euclid:2025tpw} for \hi\ IM, takes a scale cut at $k_{\rm max} = 0.15\,h\,{\rm Mpc}^{-1}$ and adopts a semi-analytic non-linear kaiser RSD model, with free non-linear terms for each bin, that also includes and accounts Fingers-of-God and BAO damping. We see that the combination of the SKAO-AA4 survey all probes will yield a bound of $\alpha_{\rm{s}} =  -0.005 \pm 0.013$ improving by a factor of two the ones on each probe alone with either the angular correlation or HI IM yielding $\sim \pm 0.025$, almost the same that we also observe for $n_{\rm s}$. We can observe a strong degeneracy between $\alpha_{\rm{s}}$ and $n_{\rm s}$. \citet{Ballardini:2016hpi,2016arXiv161205138P,Munoz:2016owz,Bermejo-Climent:2021jxf,2023MNRAS.520.2405B} showed that the combination of LSS and CMB experiments provide a powerful handle on $\alpha_{\rm{s}}$ that breaks this degeneracy. Combining all SKAO-AA4 probes with a Simons Observatory-like (CMB-SO) survey yields $\alpha_{\rm{s}} =  -0.005 \pm 0.003$ and $n_{\rm{s}} =  0.96 \pm 0.002$, improving $\alpha_{\rm{s}}$ bounds by $\sim$ 13\% from CMB alone and by $\sim$ 27\% those on $n_{\rm s}$, bringing us close to detecting a running of the primordial power spectrum if the vanilla inflation model is true. Thus, the combined constraints on $n_{\rm s}$ and $\alpha_{\rm s}$ from SKAO-AA4 and CMB-SO will also have the potential to shed light on the current inconsistency between the Planck 2018 results (TT, TE, EE + lowE + lensing), which give $n_{\rm s} = 0.9649 \pm 0.0042$ and $\alpha_{\rm s} = -0.0045 \pm 0.0067$ \citep{2020A&A...641A..10P}, and ACT~DR6 \citep{ACT:2025fju,ACT:2025tim} combined with DESI BAO \citep{DESI:2024uvr} results, which give $n_{\rm s} = 0.9743 \pm 0.0034$ and $\alpha_{\rm s} = 0.0062 \pm 0.0052$. The main advantage of SKAO compared to Stage-IV galaxy surveys is its extended redshift leverage. Increasing the redshift reach is the most effective way to increase the number of accessible linear modes. This is due to both the increased volume and the suppression of the non-linear clustering signal at higher $z$. Furthermore, as shown by \citet{2023MNRAS.520.2405B}, \hi\ IM has a $\alpha_\mathrm{s}$-$n_\mathrm{s}$ degeneracy angle that is more distinct from that of the CMB than Stage-IV galaxy surveys.

\subsection{Primordial non-Gaussianity}\label{sec:PNG}

Primordial non-Gaussianities (PNG) offer a unique window into the physics of inflation and its possible extensions. While the simplest single-field slow-roll inflationary models predict nearly Gaussian primordial perturbations, more complex scenarios can give rise to departures from Gaussianity. The most direct probe of PNG is a non vanishing bispectrum of the primordial perturbations  implicitly defined from the three-point correlation function
 \begin{equation}
 \langle\Phi(\bk_1)\Phi(\bk_2)\Phi(\bk_3)\rangle=(2\pi)^3\delta_{\rm D}(\bk_1+\bk_2+\bk_3)B_\Phi(\bk_1,\bk_2,\bk_3).
  \end{equation}
The bispectrum $B_\Phi$ can generally be written as $B(k_1,k_2,k_3)=\fnl F(k_1,k_2,k_3)$, where the shape function $F$ encodes the dependence on the triangle configurations formed by the three Fourier modes $\bk_i$, and it's amplitude is parametrised by the dimensionless parameter $\fnl$. Theoretically motivated shapes can be generated by violating the conditions of the standard inflation, with the most studied shapes being the local \citep{Salopek1990,Gangui1993,Verde1999,Komatsu2001}, equilateral \citep{Creminelli2005}, and orthogonal\footnote{Here we consider the template appropriate for LSS studies.} \citep{Senatore2009}. They are associated respectively with multi-field dynamics, non-canonical kinetic terms, and specific higher-derivative operators in the effective field theory of inflation. 
Measuring $\fnl$ with high precision has emerged as one of the most powerful probes of the early Universe \citep{2022arXiv220308128A}. Detection of $\fnl$ of different shapes would enable us to distinguish between different inflationary mechanisms and origins of primordial non-Gaussianity.

Currently, the tightest constraints on PNG come from measurements of the CMB bispectrum \citep{2020A&A...641A...9P}, which indicate that the primordial perturbation field is consistent with weak non-Gaussianity. Consequently, most of the information accessible from the CMB regarding primordial bispectra has already been extracted. LSS surveys, however, offer a powerful complementary avenue to further constrain PNG, as a three-dimensional tracer of the matter distribution (e.g. \hi) contains vastly more bispectrum modes than the two-dimensional CMB field \citep{Karagiannis2018,Meerburg_2019}. However, this additional information reside in the mildly or fully non-linear regime \citep{Jung:2022rtn}, where the gravitational evolution of structures generates its own non-Gaussianity, masking the primordial signal. 
Accurately disentangling these components requires high-precision theoretical modelling that consistently incorporates non-linear biasing and relativistic effects \citep{Maartens:2020jzf}. Such relativistic contributions include Doppler, gravitational-potential, and volume distortion terms in the number-counts counts and will be further discussed in the following section.

In the case of local PNG, LSS also offers a complementary approach to the bispectrum of dark matter tracers as it is itself sensitive to primordial non-Gaussianity. A distinctive signature arises in the power spectrum, because the coupling between large and small scale modes induced by the non-linearity in the primordial gravitational potential. It alters the connection between the dark matter tracer density field and the underlying matter perturbations. In particular, a scale dependent bias is induced \citep{Matarrese:2008nc,Dalal:2007cu}, i.e.,
\begin{equation}
    b = b_L + 3(b_L - 1) \fnl^{\text{local}} \frac{\Omega_M H_0^2 1.27 \, \delta_c}{T(k)D(k)\, k^2},
\end{equation}
where $b_L$ is the linear clustering bias, $\Omega_M$ the matter density parameter, $H_0$ is the Hubble constant, $\delta_c \approx 1.69$ is the critical overdensity for spherical collapse, $T(k)$ the transfer function, and $D(k)$ the growth. The factor $1.27$ arises from the fact that the $\fnl$ parameters are defined at the redshift of the CMB \citep{2015MNRAS.448.1035C} and this has been the convention using radio tracers of Dark matter, such as \hi\ IM and radio continuum galaxies one can put bounds on local-type PNG \citep{2015ApJ...814..145A}. This way one can directly compare with the error bars from CMB constraints. Due to the induced scale dependence, local-type PNG is the most promising target for LSS analyses. Galaxy surveys have already attempted to constrain local-type PNG using either the power spectrum alone or joint power spectrum–bispectrum analyses, although they have not yet reached the precision of CMB constraints (see, e.g., \citealp{Chaussidon:DESIpng,CagliariBarberi:eBOSSpng}). 


\subsubsection*{Constraints from the scale-dependent bias in the 2D power spectrum}

As we saw, the clustering bias gains a scale dependence in the presence of local PNG. Hence one needs ever larger surveys in volume to reach scales where the effects becomes relevant. Not only larger galaxy surveys, but also \hi\ intensity mapping promises to tighten these bounds significantly: by accessing vast cosmic volumes over a wide redshift range. It aims to achieve uncertainties on local-type PNG at the level of order unity, which is the threshold required to discriminate between different inflationary scenarios \citep{2017PhRvD..95l3507D}. This unfortunately cannot be achieved easily due to the shear volume required and future surveys, including the SKAO will not reach this limit \citep{2015ApJ...814..145A}. This is fundamentally due to cosmic variance.

A more promising way to tighten \hi\ IM constraints, particularly those from the power spectrum, is to beat cosmic variance through a multi-tracer analysis, which leverages the complementary clustering of tracers with different bias properties. Although the two-point statistics is sensitive to this effect, the PNG signal scales as $k^{-2}$, so its imprint remains confined to the largest scales. These are nevertheless affected by cosmic variance due to the limited number of independent modes available. It has been shown that with a multi-tracer approach it is possible to cancel cosmic variance, thanks to the combination of independent biased tracers of the dark matter distribution, thereby achieving a significant gain in constraining power on large-scale effects \citep{Seljak:2008mt}. There are several attempts to use multiple radio dark matter tracers, often in conjugation with optical surveys. It is worth mentioning \citet{2014MNRAS.442.2511F} and \citet{2020MNRAS.492.1513G} where they look at different radio continuum population as independent tracers of dark matter and forecast $\sigma\left(\fnl^{\text{local}}\right)\sim 4-6$. While an improvement on the forecasts from single tracers it is still far from a desired threshold of $\sigma\left(\fnl^{\text{local}}\right)< 1$.

\begin{figure}
\centering
\includegraphics[width=0.45\textwidth]{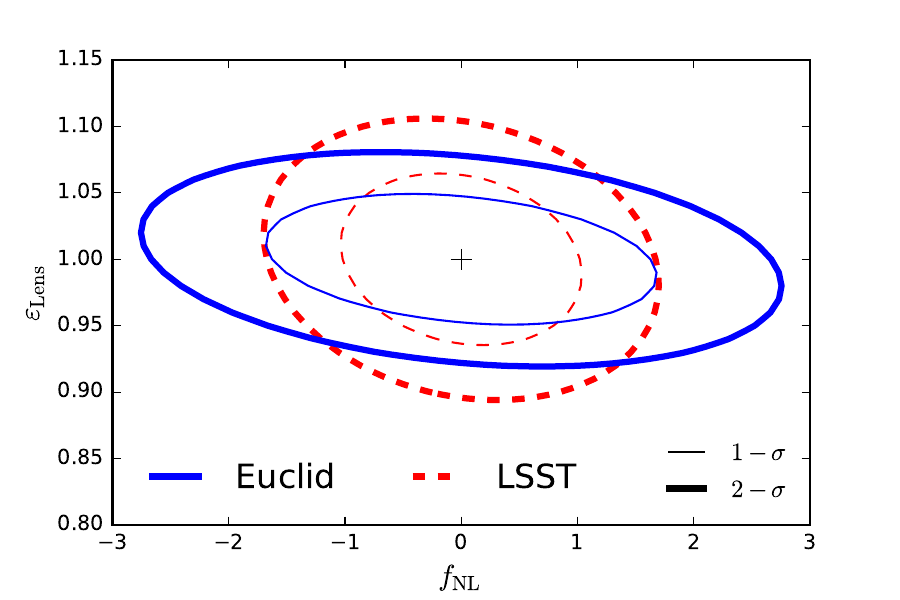}
\includegraphics[width=0.45\textwidth]{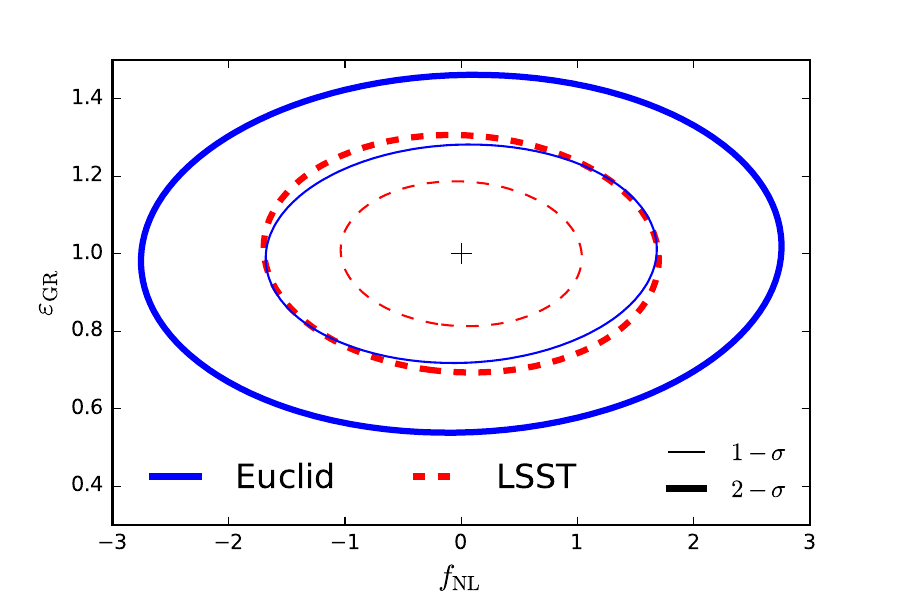}
\caption{The $1\sigma$ (thin) and $2\sigma$ (thick) contours for the forecasted marginal errors on $\fnl^{\text{local}}$ and Lensing ({\it left}), and GR effects ({\it right}) using the multi-tracer technique from \hi\ intensity mapping with the band 1 wide survey in combination with Euclid-like photometric sample (solid blue line) and LSST-like photometric sample (dashed red line). These forecasts assume Case 2 as presented in Table~\ref{tab:sigfNLGR}.}
\label{fig:MT:contour_fnl_eps}
\end{figure}

This is better achieved with \hi\ IM in conjugation with photometric galaxy surveys. Early examples \citep{2015ApJ...812L..22F,2015PhRvD..92f3525A} do show that one can reach the desired constraining power with \hi\ IM with SKAO and photometric galaxy surveys, even when marginalising over the clustering biases. Furthermore, considerable improvements in the state of the art can be achieved with MeerKAT, an SKAO precursor \citep{2017MNRAS.466.2780F}. Here we will summarise updated results for the AA4 configuration. But one should note that in the case of the scale dependent bias other effects, the so-called {\it GR effects} \citep{2011PhRvD..84d3516C,2011PhRvD..84f3505B}, introduce corrections to the tracers' transfer function at the same typical scales as $\fnl^{\text{local}}$ does. In order not to bias the estimation of PNG we have to consider these ultra-large scale effects \citep{2015MNRAS.451L..80C}. Therefore in Figure \ref{fig:MT:contour_fnl_eps} we show marginal contour plots for both $\fnl^{\text{local}}$ and the amplitude of the magnification lensing contribution and all the other GR effects, assuming an overlapping area of $10,000\deg^2$ for a Euclid-like and $14,000\deg^2$ for an LSST-like photometric surveys. These results are marginalised over the standard cosmological parameters and the clustering biases in each redshift bin. We also summarise the marginal errors for these parameters in Table~\ref{tab:sigfNLGR}. Each line in the table considers a different set of cosmological parameters of top of the standard cosmological parameters: Case 1 -- $\fnl^{\text{local}}$  only; Case 2 -- $\fnl^{\text{local}}$, the amplitudes of the Lensing contribution and all the other GR effects together; Case 3 -- $\fnl^{\text{local}}$, the amplitudes of all the individual GR effects contributions. Note that all of the $\epsilon$ parameters have a fiducial value of $\epsilon=1$. In the table, one can clearly see that one can attain $\sigma\left(\fnl^{\text{local}}\right)<1$ while providing a detection of some GR effects such as the Doppler and Lensing terms.

\begin{table*}[!t]
\caption{\label{tab:sigfNLGR} Marginal errors on $\fnl^{\text{local}}$, lensing ($\varepsilon_{\rm Lens}$) and GR effects ($\varepsilon_{\rm GR}$), which include the Doppler term ($\varepsilon_{\rm Doppler}$), Time Delay ($\varepsilon_{\rm TD}$), Sachs-Wolfe ($\varepsilon_{\rm SW}$) and Integrated Sachs-Wolfe $(\varepsilon_{\rm ISW})$, using the MT technique with \hi\ IM with the SKAO band1 wide-survey together with Euclid-like and LSST-like surveys.}
\centering
\begin{tabular}{|l|c|c|c|c|c|c|c|}
\hline
Synergy &$\sigma\left(\fnl^{\text{local}}\right)$ & $\sigma(\varepsilon_{\rm Lens})$ & $\sigma(\varepsilon_{\rm GR})$ &$\sigma(\varepsilon_{\rm Doppler})$ & $\sigma(\varepsilon_{\rm TD})$ & $\sigma(\varepsilon_{\rm SW})$& $\sigma(\varepsilon_{\rm ISW})$\\
\hline
SKAO \hi\ IM &1.1&-&-&-&-&-&-\\
$\times$  &1.1&0.033&0.19&-&-&-&-\\
Euclid-like &1.3&0.033&-&0.19&5.3&5.5&16\\
\hline
SKAO \hi\ IM  &0.67&-&-&-&-&-&-\\
$\times$ &0.68&0.043&0.12&-&-&-&-\\
LSST-like &0.96&0.043&-&0.13&5.7&4.0&7.5\\
\hline
\end{tabular}
\end{table*}

Notwithstanding, one can also attain good results with spectroscopic surveys. The forecasts presented in \cite{barberi2024} were updated in order to match the specifics of the AA4 configuration. This forecast investigates the synergy with a \textit{Euclid}-like spectroscopic galaxy survey targeting emission line galaxies in the redshift interval $0.9<z<1.8$ and covering $\approx1/3$ of the sky (with different degrees of overlap with the SKAO \hi\ intensity mapping survey). On the radio side, various systematics were taken into account, including the damping at small scales due to the beam of the SKAO dishes, a large-scale radial damping introduced to mimic signal loss from foreground cleaning, and thermal noise associated with the system temperature of the antennas. In order to account for the unavoidable parameter degeneracies, the authors performed a multi-parametric MCMC analysis to estimate the uncertainty on $\fnl^{\text{local}}$, allowing as free parameters not only $\fnl^{\text{local}}$ itself, but also the primordial spectral index $n_{\rm s}$ and the galaxy and \hi\ bias parameters, with one bias parameter per tracer for each redshift bin.
While the redshift range probed by a spectroscopic galaxy survey is more limited than that accessible with radio observations, the gain in constraining power remains significant. The foreseen constraints, however, remain far from the $\sigma\left(\fnl^{\text{local}}\right)\lesssim1$ threshold and not particularly competitive even when compared to a \hi\ single-tracer analysis but on a broader redshift coverage. If such an extended redshift coverage could also be achieved with a multi-tracer approach, the resulting constraints would become remarkable. This condition could be met by optimising the galaxy sample used in the analysis. In \cite{barberi2024}, it was proposed to exploit additional galaxy samples, distinct from the target one, in order to push the observations up to $z\leq4.4$. Alternatively, the use of \textit{DESI}-like samples of Luminous Red Galaxies and Quasars has been investigated to cover the redshift range between $0.4$ and $3.1$.
\begin{figure}
    \centering
    \includegraphics[width=0.47\linewidth]{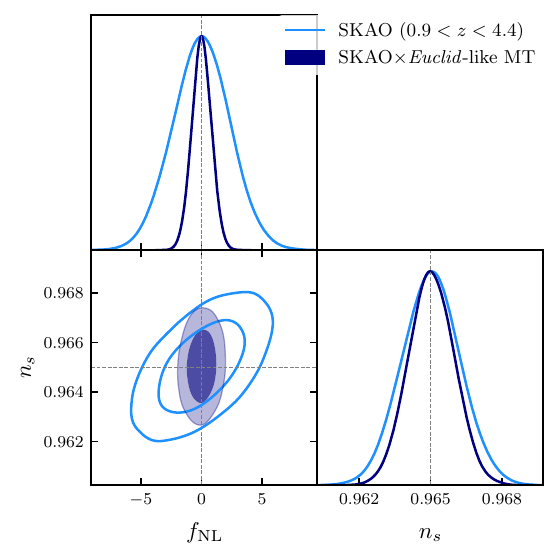}
    \hfill
    \includegraphics[width=0.47\linewidth]{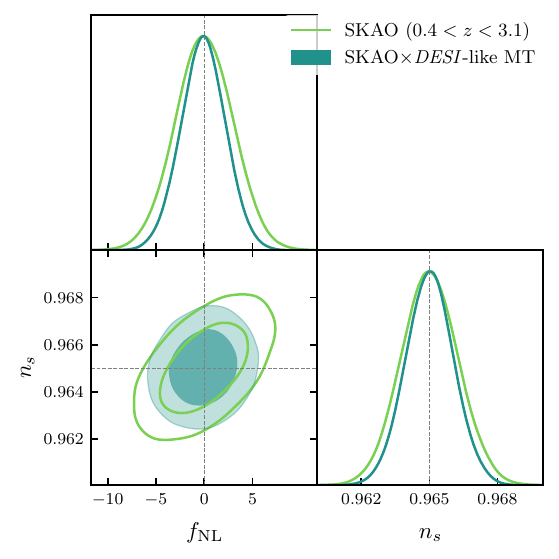}
    \caption{Corner plots showing the results of the MCMC analysis, marginalised over \hi\ and galaxy bias parameters to highlight the cosmological ones. The left panel compares the SKAO single-tracer result in the redshift range $0.9<z<4.4$ (empty contours) and the multi-tracer result obtained when combining SKAO with an extended \textit{Euclid}-like spectroscopic sample (filled contours). Similarly, on the right the \hi\ auto-correlation is compared with the multi-tracer case in the corresponding redshift range but for the combination with \textit{DESI}-like samples.}
    \label{fig:fnl_MT_MBS}
\end{figure}
The results, with a focus on the cosmological parameters, are shown in Fig.~\ref{fig:fnl_MT_MBS}. In particular, the left panel refers to the forecast performed for the SKAO single-tracer case (light-blue lines) or the multi-tracer analysis in combination with the extended \textit{Euclid}-like sample (blue filled contours). Similarly, on the right, the results are shown for the single-tracer and multi-tracer analyses but in combination with the \textit{DESI}-like samples. The wide redshift range covered in the first case extending to higher redshifts, where the signal of primordial non-Gaussianity is stronger, makes this configuration more constraining: the degeneracy between $\fnl^{\text{local}}$ and $n_{\rm s}$ is broken and the constraints from the multi-tracer technique are more than twice tighter with respect to the SKAO single-tracer analysis in the same redshift range, reaching the very promising result of $\sigma\left(\fnl^{\text{local}}\right)\approx0.9$. On the other hand, in the case of a combination with a \textit{DESI}-like sample, the redshift range probed is shifted towards lower values, resulting in a globally weaker constraining power. The improvement in the uncertainty on $\fnl^{\text{local}}$ between the SKAO single-tracer analysis and the multi-tracer case is less significant, although a breaking of the degeneracy between the parameters of the analysis is still observed. These are similar to the results obtained by \citet{2021JCAP...11..010V} where the authors perform a multi-tracer forecast between \hi\ IM and Euclid-like and DESI-like spectroscopic surveys. Finally, it will also be possible to reach $\sigma\left(\fnl^{\text{local}}\right)\sim 1$ using synergies with H$\alpha$ intensity mapping with a SPHEREx-like experiment \citep{2018MNRAS.479.3490F}.

\subsubsection*{Information from the Bispectrum}

Despite recent advances using perturbative and effective-field-theory approaches \citep{Castorina:2019wmr,Mueller:2021jbt,Cabass:2022wjy,Cabass:2022ymb,DAmico:2022gki,Chaussidon:DESIpng,CagliariBarberi:eBOSSpng}, current optical surveys still provide weaker constraints than those coming from the CMB. The primary limitation arises from the difficulty of extending analytical models beyond the perturbative regime, which is essential to exploit the full wealth of PNG information encoded in the LSS bispectrum, as well as the cosmic variance affecting the large scales of the power spectrum \citep{2015ApJ...814..145A}. 

A promising observable to overcome these limitations is \hi\ IM. Unlike traditional galaxy surveys, intensity mapping does not rely on detecting individual galaxies, enabling efficient coverage of large sky areas and redshift ranges. This makes \hi\ IM an exceptionally powerful probe of primordial non-Gaussianity.  The power spectrum of \hi\ has already been identified as a sensitive observable \citep{Camera2013,Xu2014,2015ApJ...812L..22F,2017MNRAS.466.2780F, Ballardini:2019wxj}. Moreover, the \hi\ bispectrum provides a complementary and more direct probe of primordial non-Gaussian signatures, including non-local shapes such as equilateral and orthogonal configurations that are less accessible to power-spectrum analyses, while also offering competitive constraints on the local type \citep{Karagiannis:2019jjx}. In this case, interferometric surveys outperform single-dish experiments, where the bispectrum signal is strongest \citep{Karagiannis:2020dpq}. We note that redshift uncertainties are included in our modelling. However, \hi\ intensity mapping is intrinsically spectroscopic, yielding very small redshift errors. As a result, the associated line-of-sight damping of the bispectrum is negligible over the range of scales considered in this work. Nonetheless, in a single-dish survey like SKAO, bispectrum can have a complementary role, by reducing parameter degeneracies. 

Here we summarize some of the findings of \citet{Karagiannis:2020dpq}, while we follow their formalism to present forecasts based on the SKAO AA4 specifications for constraining PNG of the local, equilateral, and orthogonal types, using both the \hi\ power spectrum and bispectrum. The authors perform Fisher matrix forecasts derived from the joint analysis of the \hi\ power spectrum and bispectrum. The tree-level model is used for both correlators, while the analysis is restricted within the perturbative regime. For equilateral and orthogonal PNG, the scale-dependent bias correction becomes effectively scale-independent on large scales, hence, only the primordial contributions in the bispectrum retain constraining power. Foreground contamination is accounted by imposing hard cuts of $k_{\parallel,\mathrm{min}} = 0.001\,h\,\mathrm{Mpc}^{-1}$ (optimistic) and $0.005\,h\,\mathrm{Mpc}^{-1}$ (pessimistic). All results in this section follow the CMB convention for $\fnl$ normalization \citep{2015MNRAS.448.1035C}. 

\begin{table}[t]
\centerline{
\begin{tabular} { l c c c }
\noalign{\vskip 3pt}\hline\noalign{\vskip 1.5pt}\hline\noalign{\vskip 6pt}
  & \bf P & \bf B & \bf P+B \\
\noalign{\vskip 3pt}\hline\noalign{\vskip 1.5pt}\hline
\multicolumn{4}{c}{ SKAO AA4} \\ 
\noalign{\vskip 3pt}\hline\noalign{\vskip 6pt}
$\sigma\left(\fnl^{\rm local}\right)$ &  1.7 \;(3.5) & 2.5\; (4.4) & 1.48\; (2.86) \\  

$\sigma\left(\fnl^{\rm equil}\right)$ &  - &  68\; (84) & 52\; (63)  \\  

$\sigma\left(\fnl^{\rm ortho}\right)$ &  - & 20\; (23) & 18\; (21) \\ 

\noalign{\vskip 3pt}\hline\noalign{\vskip 1.5pt}\hline\noalign{\vskip 6pt}
\multicolumn{4}{c}{ SKAO AA*} \\ 
\noalign{\vskip 3pt}\hline\noalign{\vskip 6pt}
$\sigma\left(\fnl^{\rm local}\right)$ &  1.8\; (3.6) & 2.6\; (4.5) & 1.53\; (2.93)  \\  

$\sigma\left(\fnl^{\rm equil}\right)$ &  - & 69\; (86) & 53\; (64)  \\  

$\sigma\left(\fnl^{\rm ortho}\right)$ &  - & 20\; (24) & 18\; (21)  \\ 

\noalign{\vskip 3pt}\hline\noalign{\vskip 1.5pt}\hline\noalign{\vskip 5pt}
\end{tabular}

}
\caption{Forecasted 1$\sigma$ marginalized uncertainties on the three PNG shapes from the complete SKAO and AA* surveys, derived from the power spectrum, bispectrum, and their combined analysis. A foreground-imposed cut of $k_{\parallel,\mathrm{min}} = 0.001\,h\,\mathrm{Mpc}^{-1}$ is assumed; values in parentheses correspond to a less optimistic case with $k_{\parallel,\mathrm{min}} = 0.005\,h\,\mathrm{Mpc}^{-1}$. All results include \planck priors on the cosmological parameters.}\label{table:PNGresults}
\end{table}
The results for both the full SKAO (AA4) and AA* configurations are summarised in Table \ref{table:PNGresults}, considering both optimistic and pessimistic foreground cuts. For the local PNG type, the power spectrum, through the scale dependent bias, dominates the constraints, while the bispectrum—although capable of providing tight limits—plays a more complementary role in the joint forecasts. In contrast, for the equilateral and orthogonal types, the bispectrum drives the constraints, consistent with previous analyses \citep[e.g.,][]{Karagiannis2018}. The forecasts also highlight the strong sensitivity of local PNG constraints to the foreground cut, as reduced access to large scales substantially degrades performance, bringing results close to current \planck limits. Nevertheless, the combined power spectrum and bispectrum signal yields competitive constraints even for the AA* specifications, with the modified design having little impact on any of the PNG types considered here. This underscores the strong constraining power of SKAO for primordial non-Gaussianity, although it still falls short of achieving the desired precision of $\sigma\left(\fnl^{\text{local}}\right)<1$. As shown in \citet{Karagiannis:2020dpq}, single-dish surveys like SKAO remain the optimal configuration for constraining local PNG through the power spectrum, while the bispectrum provides valuable complementary information, enhancing the overall robustness of PNG constraints. Finally, contributions from Band 2 are neglected, as demonstrated in \citet{Karagiannis:2020dpq} to have minimal impact.

A promising alternative is to combine single-dish and interferometer observations, by cross-correlating large-scale single-dish modes with small-scale interferometer modes. The formed bimodal bispectrum efficiently targets squeezed triangles, which carry most of the signal for local PNG \citep{Karagiannis:2024pyx}. This synergy significantly boosts bispectrum constraints on 
$\fnl^{\rm local}$, while for carefully chosen single-dish and interferometer surveys, it could potentially reach the threshold of $\sigma\left(\fnl^{\text{local}}\right)<1$ by using the bispectrum alone. The approach can avoid foreground and instrumental noise dominated regimes, while maintaining the constraining power, making it a promising path not only for local PNG but for any signal peaking on squeezed triangles.

\subsubsection*{PNG in the 1-D distribution of fluctuations}

Signals of leading-order PNG are encoded in the primordial bispectrum from which the primordial skewness is derived. Local PNG can be probed by the characteristic scale-dependent galaxy bias and potentially analogue effects in the density-split clustering \citep{Uhlemann2018pNGPDF} and voids \citep{Chan2019voidbias}.
Equilateral and orthogonal shapes can be probed with the late-time bispectrum and skewness, which require accurate modelling of the gravitationally induced non-Gaussianity that obfuscates the primordial one. The one-point probability distribution and density-dependent clustering of matter can be modelled accurately on mildly non-linear scales and is sensitive to the total primordial skewness \citep{Uhlemann2018pNGPDF,Friedrich2020fNLPDF}. Predictions for matter can be lifted to a wide range of tracers including galaxies and neutral hydrogen \citep{Leicht2019HIPDF,Friedrich2022tracerPDF,Gould2025tracerPDFspec}. The response to equilateral PNG of the related $k$-Nearest Neighbour statistics was recently shown to be distinct from cosmological, bias and HOD parameters \citep{Coulton2024pNGkNN}. This can lead to new ways to probe PNG with one-point statistics.

\subsection{Primordial features in the power spectrum}

Primordial features in the primordial power spectrum (PPS) provide a compelling insight into the fundamental physics of the early Universe, allowing for rigorous tests of the cosmic inflation paradigm that extend beyond the simplest slow-roll scenarios \citep[see][for reviews]{Chluba:2015bqa,2022arXiv220308128A}. While single-field slow-roll inflationary models suggest a nearly scale-invariant and featureless PPS, an array of theoretical frameworks introduces variations that result in distinctive patterns or \textit{features} within the spectrum.

These primordial features may arise from various physical processes occurring during inflation. For instance, abrupt changes in the inflationary potential can lead to localized oscillations or step-like structures in the PPS \citep{Starobinsky:1992ts,Adams:2001vc,Chen:2006xjb}. Moreover, if inflation is driven by multiple interacting scalar fields rather than a single scalar field inflaton, the resulting dynamics could lead to modulations, including resonant oscillations or transient bumps and dips \citep{Chen:2008wn,Flauger:2009ab,Flauger:2010ja,Chen:2010bka,Achucarro:2010da,Chen:2011zf,Chen:2014cwa}.
The detection and characterization of these features are crucial since they encode information about high-energy physics and the underlying inflationary mechanisms, potentially unveiling interactions or particle content that are otherwise inaccessible. Moreover, departures from a simple power-law primordial spectrum, such as oscillatory or step-like features, can slightly modify the inferred values of late-time parameters like the Hubble constant ($H_0$) and the amplitude of matter fluctuations ($\sigma_8$), by changing the shape of the CMB and matter power spectra. Such models have been explored as possible ways to alleviate both the Hubble and $S_8$ tensions \citep{Hazra:2018opk,Lee:2025yah,Hazra:2022rdl}.

In recent decades, efforts to uncover primordial features have predominantly centred on measurements of CMB anisotropies, examining both the angular power spectrum \citep{2020A&A...641A..10P,Raffaelli:2026bhk} and higher-order statistics such as the bispectrum \citep{2020A&A...641A...9P}. These investigations have placed significant constraints on deviations from a power-law PPS but have yet to uncover definitive evidence of features.\footnote{Model-independent PPS reconstruction from current CMB power spectrum measurements suggest that any deviations from scale invariance must be within a few percent of the PPS amplitude over the range $0.005 \lesssim k/({\rm Mpc}^{-1}) \lesssim 0.25$ \citep{Raffaelli:2025kew}.} More recently, LSS surveys have emerged as valuable complementary probes, providing higher resolution measurements of the matter power spectrum on smaller scales \citep{Huang:2012mr,Chen:2016vvw,Ballardini:2016hpi,Beutler:2019ojk,Ballardini:2022wzu,Ballardini:2022vzh,Mergulhao:2023ukp,2023arXiv230917287B,Calderon:2025xod,Karagiannis:2026xkv}. The enhanced sensitivity of LSS data to oscillatory signals holds promise for refining constraints on primordial features and potentially revealing subtle imprints that may have eluded CMB observations.
This has motivated a dedicated effort toward accurately modelling and understanding the galaxy power spectrum in the mildly non-linear regime \citep{Vlah:2015zda,Blas:2016sfa,Vasudevan:2019ewf,Beutler:2019ojk,Ballardini:2019tuc,Chen:2020ckc,Ballardini:2024dto}.

21cm intensity mapping and radio continuum surveys planned for SKAO  will place extremely tight constraints on the parameters of feature models even without the inclusion of CMB data \citep{Chen:2016zuu,Xu:2016kwz,Meerburg:2016zdz,Ballardini:2017qwq,Bacon:2018dui,Karagiannis:2026xkv}. Beyond the statistical improvement, 21cm observations offer unique advantages: their large sky fraction and sensitivity to high redshifts make them particularly suited for probing features at large scales, while their ability to access the mildly non-linear regime at early epochs could help recover part of the information lost to bulk motions in the case of oscillatory features and localized features at small or intermediate scales. Combining the \hi\ IM power spectrum and bispectrum to constrain feature parameters yields a substantial improvement in sensitivity, as \hi\ intensity mapping naturally satisfies the observational requirements for precise bispectrum analyses \citep{Karagiannis:2026xkv}.

\section{Theoretical and observational challenges}

In practice, one has challenges that make a detection of primordial non-Gaussianity harder. In the specific case of \hi\ intensity mapping there are foreground contaminants that affect the power spectrum in scales we are most interested in in this chapter. Additionally, other ultra large-scale effects alter our summary statistics in the same way. The better understood of those are the ones with the so-called {\it GR effects} \citep{2011PhRvD..84d3516C,2011PhRvD..84f3505B}. In this section we summarize the implications of signal loss from foreground cleaning and the implications of not modelling properly the power spectra on ultra-large scales. We mainly specify the impact of these in PNG as they are the most widely studied in the literature.

\subsubsection*{Large-scale signal loss from intensity map foreground contamination}

A central challenge in 21cm intensity mapping surveys is the presence of bright astrophysical foregrounds, several orders of magnitude stronger than the cosmological \hi\ signal. These foregrounds, arising primarily from Galactic synchrotron, free-free emission and extragalactic continuum sources, are spectrally smooth compared to the 21cm signal and can in principle be separated in frequency space \citep{Ansari:2011bv,Wolz:2013wna}. In practice, however, the removal of foregrounds is non-trivial and inevitably leads to distortions of the cosmological signal.

The most widely used approaches rely on blind foreground removal techniques, such as principal component analysis (PCA) and related subspace methods \citep{Alonso:2014dhk,Cunnington:2020njn,Spinelli:2021emp}. These techniques exploit the strong frequency coherence of the dominant foregrounds and remove a subset of the most correlated spectral modes. Over the past decade, extensive tests with simulations and observational data have made these methods increasingly robust, establishing them as the standard foreground cleaning strategy for upcoming surveys, including SKAO and its pathfinders \citep{MeerKLASS:2024ypg,Carucci:2024qpm}. However, since the cosmological \hi\ signal also possesses large-scale spectral correlations, blind cleaning not only removes foregrounds but also suppresses cosmological fluctuations on the largest scales, exactly those scales most sensitive to primordial non-Gaussianity. Without sufficient correction for this \textit{signal loss}, measurements of the local-type non-Gaussianity parameter, $f_{\rm NL}$, can suffer catastrophic biases, in some cases exceeding 12$\sigma$ \citep{Cunnington:2020wdu}. 

One possible mitigation strategy is to include nuisance parameters that model the signal loss within the analysis pipeline \citep{Cunnington:2020wdu,Fonseca:2020lmi}. While this can remove the bias, it introduces strong degeneracies with the $f_{\rm NL}$ signal, dramatically inflating the resulting uncertainty. However, recent work has shown that signal loss can be reconstructed by using mock signal injection to create a foreground transfer function, which unbiases the power spectrum measurement \citep{Switzer:2015ria,Cunnington:2023jpq}. This has been shown to be relatively model-independent, meaning it can be used for parameter inference, potentially with only modest increases in statistical errors, providing realistic hope that foreground cleaning and loss correction can be combined into a robust pipeline for SKAO cosmology.

A common assumption in the forecasting literature has been to impose a sharp scale cut, removing all modes below some $k_\parallel$ threshold in order to sidestep these complications. This treatment, however, is overly pessimistic and fundamentally incomplete: signal loss is not confined to the very largest scales but affects a broad range of modes, albeit at varying levels. Moreover, by discarding these modes, such forecasts understate the potential of emerging signal-reconstruction methods that can recover much of the lost information. \citet{Cunnington:2020wdu} and \citet{Fonseca:2020lmi} present forecasts that do not impose such scale cuts, and instead assume that signal loss can be corrected and marginalised over to a level sufficient for unbiased cosmological inference. This represents a realistic and optimistic target for SKAO, given the rapid progress in both foreground cleaning algorithms \citep{Carucci:2024qpm} and signal-loss reconstruction methods \citep{Chen:2025til} that will continue to develop in the pre-SKAO era.

\subsubsection*{Degeneracies with light-cone effects}

The ultra-large scale are precisely the scales where relativistic light-cone and wide-angle projection effects also become relevant. These come from the fact that one measures angular positions and redshifts instead of real space distances. The projection between the observed space to real space in a perturbed universe gives rise to corrections to the observed density contrast which we call the GR effects. If such effects are neglected in the theoretical modelling of the observed power spectrum, they can partially
mimic the signature of local PNG and lead to a systematic shift in the inferred value of $f_{\mathrm{NL}}$ \citep{2015MNRAS.451L..80C,2021JCAP...12..004V,2025JCAP...07..063G}. This effect can be quantified by comparing a ``true'' model, which includes all relativistic (and wide-angle contributions in the case of the 3D power spectrum) to the observed power spectra, with an ``approximate'' model that retains only the standard power spectra (usual matter density and redshift-space distortion terms). Within a Fisher-matrix framework, the shift in a cosmological parameter $\vartheta_\alpha$ induced by this modelling mismatch is given by \citep{Fonseca:2020lmi}
\begin{align}
\delta \vartheta_\alpha = \sum_\beta \left(F^{-1}\right)_{\alpha\beta} \, G_{\beta \epsilon} \ \delta \vartheta_\epsilon  \, , \label{eq:parameter_shift}
\end{align}
where $F_{\alpha\beta}$ is the Fisher matrix computed using the approximate model (standard power spectrum), while $G_{\beta \epsilon}$ is the Fisher matrix of the full proper model (with  relativistic and wide-angle corrections) and, $\delta \vartheta_\epsilon$ is the shift from neglecting such corrections.  As an example, $\vartheta_\epsilon = 0$ is the ``approximate'' model and $\vartheta_\epsilon = 1$ in the ``true'' model. The shift $\delta f_{\mathrm{NL}}$ measures the displacement of the best-fit PNG amplitude caused by ignoring ultra-large-scale effects.

In \citet{2021JCAP...12..004V} and \citet{2025JCAP...07..063G} the authors look at the shift in the best fit of $\fnl^{\text{local}}$ using \hi\ IM with SKAO on its own, and together with a galaxy survey. For an \hi\ IM survey alone, such as SKAO Band~1, the induced shift in $f_{\mathrm{NL}}$ is found to be very small, at the level of $\delta f_{\mathrm{NL}} \simeq 0.1\,\sigma(\fnl^{\text{local}})$. This is primarily due to the fact that the dominant integrated relativistic contribution, lensing magnification, vanishes at first order in intensity mapping, while the remaining non-integrated corrections are subdominant on the largest scales. Consequently, SKAO-only analyses are essentially unaffected in terms of the best-fit value of $\fnl^{\text{local}}$. This can be seen as an advantage of \hi\ IM. The situation changes significantly in a multi-tracer analysis combining SKAO with a galaxy survey such as an Euclid-like one. Galaxy number counts are significantly affected at high $z$. The multi-tracer technique strongly suppresses cosmic variance on ultra-large scales, thereby enhancing sensitivity to any residual model ling error. As a result, when relativistic and wide-angle effects are neglected, the inferred value of
$\fnl^{\text{local}}$ in a joint SKAO $\otimes$ Euclid-like analysis is shifted by $\delta f_{\mathrm{NL}} \simeq 2\,\sigma(\fnl^{\text{local}})$. This highlights that, while relativistic effects induce only a negligible shift in $\fnl^{\text{local}}$ for SKAO alone, they lead to a significant displacement of the best-fit value in multi-tracer analyses. Accurate modelling of ultra-large-scale relativistic contributions is therefore essential to avoid biased measurements of primordial non-Gaussianity when combining SKAO with spectroscopic galaxy surveys. And, as already seen in Table~\ref{tab:sigfNLGR}, marginalising over the amplitudes of the GR effects does not have a severe impact in the constraining power when adding galaxy surveys to \hi\ IM.

In the case of the Bispectrum, the light cone effects also have a considerable impact in the estimates of the different $\fnl$ parameters. A full observed-space description including wide-angle and GR effects relevant to forthcoming all-sky surveys, was obtained in \citet{Addis:2024zhw} and \citet{Addis:2025rre} for the multipoles of the 3D power spectrum. The relativistic terms have non-negligible contribution to the observed bispectrum in the same scales as PNG and can mimic parts of it unless they are modelled consistently. As an example, a coupling between large- and small-scale modes produced by local relativistic contributions enter in the bispectrum in a way that partly overlaps with squeezed-limit PNG \citep{Maartens:2020jzf,Rossiter:2024tvi}. Hence neglecting them can lead to non-negligible parameter bias on $f_{\rm NL}$, meaning that foreground mitigation and relativistic modelling are complementary requirements for unbiased PNG inference.

In \ref{sec:PNG} one explore the complementarity between two-point and three-point statistics. This complementarity is especially relevant on ultra-large scales, where relativistic projection and wide-angle effects evolve differently with redshift from the primordial component. An analytic decomposition that cleanly separate these two sources of long-short-mode coupling exists\citep{Maartens:2020jzf,Addis:2025rre,Rossiter:2024tvi} and is crucial for unbiased results.

\section{Discussion}



In this chapter we reviewed how the SKAO can provide transformational information on what may have happened in the nascent cosmos. While there is sufficient evidence to back an inflationary early universe \citep{2020A&A...641A..10P} one is ought to provide consistent observational constraints and explore the limits of the model. Therefore, we have briefly summarised what inflation is and discussed which predictions can be probed by cosmological observations with the SKAO, namely \hi\ IM and Radio Continuum surveys. 

There are three main observable parameters which we can probe with the SKAO: the running of the power spectrum $\alpha_{\rm s}$, primordial non-Gaussianity $\fnl$ and primordials features in the power spectrum. In the case of $\alpha_{\rm s}$ adding information coming from radio tracers of Dark matter one can provide a 13\% improvement on CMB information alone. In the case of PNG one has several avenues in which the SKAO will be used constrain such parameters. One can use three-point functions of the density contrast of different radio tracer alone or in conjugation with optical surveys. For SKAO-like sky coverage, these three-point statistics ultimately benefit from formulations that include Doppler, lensing, potential, integrated line-of-sight and wide-angle contributions in the observed bispectrum. One can also use the two-point functions, either in 3D space or projected on the sphere, to probe the bias of the particular tracer to probe a PNG-induced scale dependence. Several works have looked at this but it is only synergies between surveys that can provide the threshold constraint $\sigma\left(\fnl^{\text{local}})< 1\right)$. Such constraints can only be obtained when adding information from the SKAO. Achieving such precision also relies on theoretical developments that clarify the interplay between wide-angle effects and PNG in the power spectrum and bispectrum, helping ensure that SKAO-driven improvements are not limited by modelling systematics. Finally primordial features in the primordial power spectrum are a direct probe of the fundamental physics of the early universe that only become accessible on the largest cosmological scales today, such as the ones covered by the SKAO cosmological surveys. 

This chapter looks at physics beyond $\Lambda$CDM in the early universe while \cite{Camera01.2026.SKA} focus on tests of gravity in cosmological scales. It is important to note that the scientific tests foreseen in this chapter are dependent on keeping under control all observational systematics \citep{Spinelli01.2026.SKA}.

\section*{Author List Ordering}

Authors for this chapter are ordered as: corresponding author, alphabetical tiers. Authors in the first tier contributed to the chapter by producing original results using SKAO updated specifications or writing significant part of the text with results using SKAO AA4 specifications. Authors in the second tier contributed with useful comments and constructive feedback, active supervision, and valuable inputs for discussion and small edits to the text.

\section*{Contribution}

JF overviewed the chapter. BBK and ZS contributed to the discussion on the running of the spectral index. JF, MBS, CU, DK contributed with the discussion on primordial non-Gaussianity. MB contributed with the discussion on primordial features in the power spectrum. JF, SC, SR and SG contributed with theoretical and observational challenges. All the authors contributed with discussion, edits and revision of the chapter.

\section*{Acknowledgments}

JF acknowledges support of Funda\c{c}\~{a}o para a Ci\^{e}ncia e a Tecnologia (FCT) through the Investigador FCT Contract no.\ 2020.02633.CEECIND/CP1631/CT0002, the FCT exploratory project 10.54499/2023.15069.PEX, and the research grant UID/04434/2025. BB-K acknowledges support from INAF for the project `Paving the way to radio cosmology in the SKA Observatory era: synergies between SKA pathfinders/precursors and the new generation of optical/near-infrared cosmological surveys' (CUP C54I19001050001). SCa acknowledges support from the Italian Ministry of University and Research (\textsc{mur}), PRIN 2022 `EXSKALIBUR – Euclid-Cross-SKA: Likelihood Inference Building for Universe's Research', Grant No.\ 20222BBYB9, CUP D53D2300252 0006, from the Italian Ministry of Foreign Affairs and International
Cooperation (\textsc{maeci}), Grant No.\ ZA23GR03, and from the European Union -- Next Generation EU. DK acknowledges support by \textsc{mur}, PRIN 2022 `BROWSEPOL: Beyond standaRd mOdel With coSmic microwavE background POLarization', Grant No.\ 2022EJNZ53. BRD is supported by the South African Radio Astronomy Observatory and the National Research Foundation (Grant No. 75415). ZS acknowledges support from the research projects PID2021-123012NB-C43, PID2024-159420NB-C43, the Proyecto de Investigación SAFE25003 from the Consejo Superior de Investigaciones Científicas (CSIC), and the Spanish Research Agency (Agencia Estatal de Investigaci\'on) through the Grant IFT Centro de Excelencia Severo Ochoa No CEX2020-001007-S, funded by MCIN/AEI/10.13039/501100011033.
\bibliographystyle{abbrvnat-maxbibnames4}
\bibliography{chapter}

\end{document}